\documentstyle[twoside,fleqn,espcrc2,psfig,graphics]{article}

\newcommand{\AmS}{{\protect\the\textfont2
  A\kern-.1667em\lower.5ex\hbox{M}\kern-.125emS}}

\title{The Gribov Ambiguity for Maximal Abelian and Center Gauges
in SU(2) Lattice Gauge Theory}

\input epsf

\author{ John D. Stack   and William W. Tucker
\address{Department of Physics, University of Illinois at 
Urbana-Champaign, 1110 West Green Street, Urbana, Illinois, 68101}
\thanks{This work was supported in part by the National Science 
Foundation.  The computations on the SV1 system (Texas, Austin)
were supported by NPACI.
}}
       
\begin{document}

\begin{abstract}
We present results for the fundamental string tension in SU(2) lattice
gauge theory after projection to maximal abelian and direct maximal center
gauges.  We generate 20 Gribov copies/configuration.   Abelian and center
projected string tensions slowly decrease as 
higher values of the gauge functionals are reached.
\end{abstract}

\maketitle

The suggestion that a clearer view of confinement could be attained after
gauge-fixing was first made by 't Hooft, who also introduced the maximal
abelian gauge (MAG) \cite{thooft}.  
For an $SU(2)$ gauge theory, this gauge attempts to
suppress the `charged' gauge fields $A_{\mu}^{1},A_{\mu}^{2}$, and maximize
the `abelian' field $A_{\mu}^{3}$.  In the continuum this is done by
finding the minimum over the group of gauge transformations of the functional
\begin{equation}
G^{c}_{mag}= \frac{1}{2}\sum_{\mu}\int \left ( (A^{1}_{\mu})^{2}
+(A^{2}_{\mu})^{2} \right )d^{4}x.
\label{contgf}
\end{equation}
Minimizing $G_{mag}^{c}$ is equivalent in the limit of small lattice spacing
to maximizing the lattice functional
\begin{equation}
G_{mag}^{l}=\frac{1}{2N_{link}}\sum_{x,\mu}
{\rm tr}\left[U^{\dagger}_{\mu}(x)\sigma_{3}U_{\mu}(x)
\sigma_{3}\right],
\label{latmag}
\end{equation}
where $N_{link}$ is the number of links on the lattice. The normalization is
such that $G_{mag}^{l}$ 
would be unity if every link were abelian.  

Another gauge which has been widely studied in $SU(2)$ lattice gauge theory
is the direct maximal center gauge (DCG) \cite{greensite1}, 
which seeks the maximum of the functional
\begin{equation}
G_{dcg}^{l}=\frac{1}{4N_{link}}\sum_{x,\mu}
({\rm tr}\left[U_{\mu}(x)\right])^{2}.
\label{latdcg}
\end{equation}
$G_{dcg}^{l}$ is normalized so that it would be unity if every link were
a member of the center $Z(2)$  of $SU(2)$.

\section{Gribov Ambiguities}

Both the maximal abelian and maximal center gauges are subject to the
Gribov ambiguity.  This means that on a given lattice configuration,
the functionals $G^{l}_{mag}$ and $G^{l}_{dcg}$ have many local maxima.
The numerical application of a gauge-fixing algorithm will eventually
put the configuration
in the near vicinity of some local maximum.  If before the application of
the gauge-fixing algorithm, a  random gauge transformation
is applied, then upon gauge-fixing, the  configuration 
in general approaches a different local maximum.  A large number of 
so-called `Gribov
copies'  of the  original configuration can be generated in this way. A
gauge invariant quantity will of course take the same value on any one of these
copies.  However, if the gauge invariant quantity is estimated by using
approximate or `projected' links, the value obtained will depend on which
Gribov copy is utilized.

For the MAG, the link can be factored as
\begin{equation}
U_{\mu}(x)= U_{\mu}^{u1}(x)\cdot U_{\mu}^{\perp}(x),
\end{equation}
where the abelian link $U_{\mu}^{u1}$ is a diagonal matrix \cite{stack}.
Abelian projection  means approximating the full $SU(2)$ 
link $U_{\mu}$ by its abelian part, $U_{\mu}^{u1}$.  

For the DCG, the link is factored as
\begin{equation}
U_{\mu}(x)= Z_{\mu}(x)\cdot U_{\mu}^{\prime}(x),
\end{equation}
where $Z_{\mu}(x)={\rm sign}({\rm tr}[U_{\mu}(x)])$.  Center 
projection means replacing the full $SU(2)$ link by its center part, $Z_{\mu}$.
\begin{table*}[htb]
\setlength{\tabcolsep}{3.5pc}
\newlength{\digitwidth} \settowidth{\digitwidth}{\rm 0}
\catcode`?=\active \def?{\kern\digitwidth}
\caption{Functional and string tension values for 
MAG}

\label{tab1}
\begin{tabular*}{\textwidth}{@{}l@{\extracolsep{\fill}}rrrr}
\hline
                 & \multicolumn{1}{r}{$N_{copy}$}
                 & \multicolumn{1}{r}{$G_{mag}^{l}$}
                 & \multicolumn{1}{r}{$\sigma_{U(1)}$}
                 & \multicolumn{1}{r}{$\sigma_{mon}$}  \\
\hline
&1 & 0.7493(1) &   0.038(1)    &  0.0329(2) \\
&10 & 0.7498(1) &   0.034(1)    &  0.0299(4) \\
&20 & 0.7499(1) &   0.033(1)    &  0.0292(5)\\
\hline
\end{tabular*}
\end{table*}

\begin{table*}[htb]
\setlength{\tabcolsep}{3.5pc}
\catcode`?=\active \def?{\kern\digitwidth}
\caption{Functional values, string tension, and P-vortex density for  DCG}
\label{tab2}
\begin{tabular*}{\textwidth}{@{}l@{\extracolsep{\fill}}rrrr}
\hline
                 & \multicolumn{1}{r}{$N_{copy}$}
                 & \multicolumn{1}{r}{$G_{dcg}^{l}$}
                 & \multicolumn{1}{r}{$\sigma_{Z(2)}$}
                 & \multicolumn{1}{r}{$\rho_{Z(2)}$}  \\
\hline
&1 & 0.7917(1) &   0.0395(5)    &  0.0315(1) \\
&10 & 0.7925(1) &   0.0351(4)    &  0.0299(2) \\
&20 & 0.7927(1) &   0.0343(4)    &  0.0296(1)\\
\hline
\end{tabular*}
\end{table*}

The effect of the Gribov ambiguity on the fundamental string tension has
been studied for the maximal abelian gauge in \cite{hart,bali1},
and for the maximal center gauge in \cite{bornyakov1,bornyakov2,greensite2}.
Since numerically finding the 
global 
maximum of the gauge functional is not really feasible,
the procedure in practice has been to generate  several copies/configuration, 
and
determine how  string tension estimates vary as 
higher values of the gauge functionals
are reached.  In all cases, it is found that
higher values of the
gauge functionals $G_{mag}^{l}$ and $G_{dcg}^{l}$ are correlated with
lower
estimates for the fundamental string tension.  

\vspace{-0.3cm}

\section{Stopping Criteria}
We have applied overrelaxation methods \cite{mandula} 
to finding local maxima of
the functionals $G_{mag}^{l}$ and $G_{dcg}^{l}$.  The
iteration of the algorithm is stopped when a certain criterion is
met, implying that the configuration is exceedingly close to a local maximum.
For the MAG, a local maximum of $G_{mag}^{l}$ implies that
\begin{eqnarray*}
X_{i}(x)&\equiv& \frac{1}{2}\left\{\sum_{\mu}{\rm tr}
\left[\sigma_{i}U_{\mu}(x)\sigma_{3}
U^{\dagger}_{\mu}(x)\right]\right.\\
& &\left.+\sum_{\mu}{\rm tr}\left[\sigma_{i}U^{\dagger}_{\mu}(x-\hat{\mu})
\sigma_{3}U_{\mu}(x-\hat{\mu})\right]\right\}\\
& & =0,\;\;i=1,2.
\label{Xdef}
\end{eqnarray*}
Defining the quantity
\begin{equation}
R_{mag} = \frac{1}{N_{site}}\sum_{x}\sum_{i=1}^{2}\left[
|X_{i}(x)|^{2}\right],
\end{equation}
we demand that $R_{mag} \leq 10^{-10}$.

For the DCG, a local maximum of $G_{dcg}^{l}$ implies that
\begin{eqnarray*}
X_{i}(x)&\equiv& \frac{1}{4}\left\{
\sum_{\mu}{\rm tr}\left[\sigma_{i}U_{\mu}(x)\right]
\cdot {\rm tr}\left[U_{\mu}(x)\right]\right.\\
& & \left.-\sum_{\mu}{\rm tr}\left[\sigma_{i}U_{\mu}(x-\hat{\mu})\right]
\cdot {\rm tr}\left[U_{\mu}(x-\hat{\mu})\right]\right\}\\
& & =0,\;\;i=1,2,3.
\end{eqnarray*}

Defining the quantity
\begin{equation}
R_{dcg} = \frac{1}{N_{site}}\sum_{x}\sum_{i=1}^{3}\left[
|X_{i}(x)|^{2}\right],
\end{equation}
we demand that $R_{dcg} \leq 10^{-12}$.

\section{Results}

Our calculations used a $24^{4}$ lattice at $\beta=2.50$.  The
results presented here are for 30 widely separated configurations.  We 
gathered 20 Gribov copies/configuration for both MAG and DCG,  where
each copy satisfied the conditions of the previous section.  We measured 
three types
of projected Wilson loops $W(R,T)$, for $R=2-8,T=2-10$.  For the MAG
we measured $U(1)$ and monopole Wilson loops. The monopole Wilson loops
are calculated from the
magnetic currents of monopoles.  
The magnetic current is found by locating edges of Dirac sheets
 \cite{degrand,stack}.
For the DCG, we measured $Z(2)$ Wilson loops, which can be expressed
in terms of the number of  P-vortices piercing
the loop \cite{greensite1}.  
The density of P-vortices, which is the same as the density of
negative $Z(2)$ plaquettes,
was also monitored.

The variation of results with the number 
of  Gribov  copies was handled exactly as
in \cite{bali1}.   All possible sets of size $N_{copy}\le 20$  
were formed for each configuration, and on each set 
the copy with highest gauge functional was chosen. 
Wilson loops were computed, averaged over the sets,
then averaged over configurations.  The resulting average
Wilson loops, indexed by $N_{copy}$,
were used 
to extract potentials $V(R)$, by fitting $-\ln(W(R,T))$ to a
straight line in $T$.  Finally, string tensions were obtained by
a linear-plus-Coulomb fit to $V(R)$ over the range $R=2-8$.

The gauge functionals, string tension values, and P-vortex densities are
tabulated in Tables \ref{tab1} and \ref{tab2}.
From the tables, it is seen that a quite small relative increase in the gauge
functionals  corresponds to a  much larger relative decrease in
the string tensions.
Comparing Tables \ref{tab1} and
\ref{tab2}, it is interesting to note the tendency of the monopole string
tension to lie close to the density of P-vortices.

\begin{figure}[htb]
\vspace{18pt}
\psfig{file=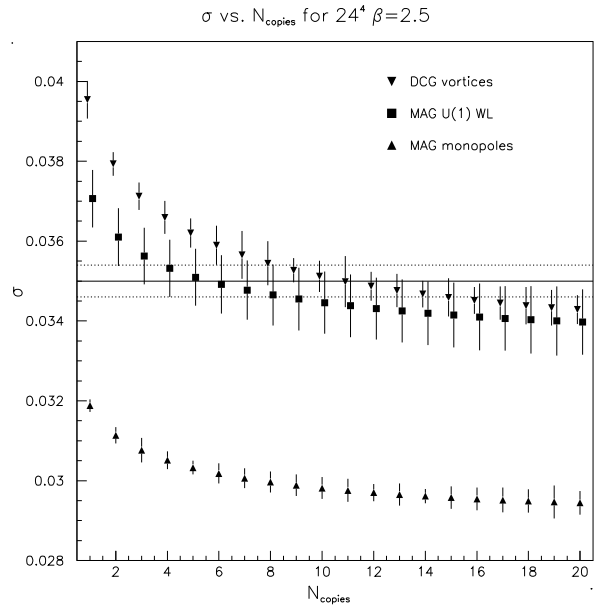,height=8.cm}
\vspace{-1.0cm}
\caption{The fundamental string tension for MAG ($U(1)$ and monopoles)
and DCG \mbox{(P-vortices)} vs. number of Gribov copies.}
\label{fig1}
\end{figure}

Figure \ref{fig1} shows the string tension estimates vs $N_{copy}$, along
with the full
$SU(2)$ string tension (0.350(4)) from \cite{bali2}.
As the plot makes clear, taking one copy/configuration overestimates
the string tensions, by an amount well outside of error bars. The results for
$\sigma_{U(1)}$ and $\sigma_{Z(2)}$ are 
much closer to the full $SU(2)$ value
for $N_{copy}=20$ than they are for $N_{copy}=1$.  However,
despite the apparent flattening out of the data
as $N_{copy}$ increases, a very gradual
descent to much smaller string tensions could still occur 
for $N_{copy}\gg 20$.  That this 
is what happens for $\sigma_{Z(2)}$ is suggested by the results of 
\cite{bornyakov2},
which finds significantly lower 
values of $\sigma_{Z(2)}$
than we report here, as well as higher functional values.

\end{document}